%% file: pssb-Kondo_v2.tex
\documentclass[pss]{wiley2sp} 
\usepackage{amsmath}
\usepackage{amssymb}

\input{ourmacros}

\begin{document}

\title{Kondo effect in the Kohn-Sham conductance of multiple levels quantum dots}

\titlerunning{Kondo effect in Kohn-Sham conductances}

\author{%
  Gianluca Stefanucci\textsuperscript{\Ast,\textsf{\bfseries 
  1},\textsf{\bfseries 2},\textsf{\bfseries 3}} and
  Stefan Kurth\textsuperscript{\textsf{\bfseries 4},\textsf{\bfseries 
  5},\textsf{\bfseries 3}}   }

\authorrunning{G. Stefanucci et al.}

\mail{e-mail
  \textsf{gianluca.stefanucci@roma2.infn.it}}

\institute{%
  \textsuperscript{1}\,Dipartimento di Fisica, Universit\`{a} di Roma Tor Vergata,
Via della Ricerca Scientifica 1, 00133 Rome, Italy\\
  \textsuperscript{2}\,INFN, Laboratori Nazionali di Frascati, Via E. Fermi 40, 00044 Frascati, 
Italy\\
  \textsuperscript{3}\,European Theoretical Spectroscopy Facility 
  (ETSF)\\
   \textsuperscript{4}\,Nano-Bio Spectroscopy Group, 
Dpto. de F\'{i}sica de Materiales, 
Universidad del Pa\'{i}s Vasco UPV/EHU, 
Avenida Tolosa 72, E-20018 San Sebasti\'{a}n, Spain\\ 
  \textsuperscript{5}\,
  IKERBASQUE, Basque Foundation for Science, E-48011 Bilbao, Spain
  }

\received{XXXX, revised XXXX, accepted XXXX} 
\published{XXXX} 


\abstract{%
\abstcol{%

At zero temperature, the Landauer formalism combined with static density 
functional theory is able to correctly reproduce the Kondo plateau in the 
conductance of the Anderson impurity model provided that an 
exchange-correlation potential is used which correctly exhibits steps at 
integer occupation.  

Here we extend this recent finding to multi-level quantum dots described by the 
constant-interaction model. We derive the exact exchange-correlation potential 
in this model for the isolated dot and deduce an accurate approximation for the 
case when the dot is weakly coupled to two leads. We show that at zero 
temperature and 
}{%
for non-degenerate levels in the dot we correctly 
obtain the conductance plateau for any odd 
number of electrons on the dot. We also analyze 
the case when some of the levels of the dot are degenerate and again 
obtain good qualitative agreement with results obtained with 
alternative methods. 

As in the case of a single level, for temperatures larger than the Kondo 
temperature, the Kohn-Sham conductance fails to reproduce the typical 
Coulomb blockade peaks. This is attributed to {\em dynamical} 
exchange-correlation corrections to the conductance originating from 
time-dependent density functional theory.
}}

\maketitle   

\section{Introduction}

Common wisdom has it that Density Functional Theory (DFT) is not able to 
describe strongly correlated systems although the fundamental theorems of 
DFT apply to these systems as well. In fact, the difficulties of dealing with 
strong correlations are not inherent to  DFT but to the DFT 
approximations.
In the last years there has been considerable progress in 
understanding which features a DFT approximation should have
in order to capture strong correlation 
effects~\cite{loc.2002,lsoc.2003,mscy.2009,mgg.2012,mmcrg.2013,xctk.2012,lyb.2012,kk.2013,v.2008,kskvg.2010,hk.2012,efrm.2012,nrvl.2013}, 
and the derivative discontinuity of the 
exchange-correlation energy functional \cite{pplb.1982} 
has emerged as one of the key properties.
In the context of quantum transport the derivative discontinuity 
turned out to be crucial to reproduce the conductance plateau~\cite{sk.2011,blb.2012,tse.2012} due to 
the  Kondo effect, a hallmark of 
strong correlations. In this paper we extend the analysis of 
Ref.~\cite{sk.2011} to multiple levels and finite temperature. In 
particular we investigate the reliability of the exact Kohn-Sham (KS) 
conductance in multi-level quantum-dot systems. The paper is organized as follows. 
In Section \ref{cimsec} we introduce the 
Constant 
Interaction Model (CIM) for the description of a multi-level quantum 
dot, and derive the exact xc potential of the isolated 
dot as well as an approximate, but accurate, xc potential 
for the dot coupled to leads.
In Section \ref{denscondsec} we discuss the self-consistent DFT 
equations of 
the CIM  and how to calculate the KS conductance at  
finite bias and temperature. Results on different quantum dots are 
presented in Section \ref{ressec}. Our main findings are that (i) the 
zero-temperature KS conductance correctly exhibits plateaus of height 
$nG_{0}$ where $G_{0}=2e^{2}/h$ is the quantum of conductance and 
$n=1,\ldots,g$, with $g$ the degeneracy of the levels,  
(ii) if only $g'<g$ degenerate levels are coupled to the leads then 
the height of the plateaus does not exceed $g'G_{0}$ and the length of 
the plateau of height $g'G_{0}$ is $[2(g-g')+1]$ times the charging energy. 
and (iii) the finite temperature KS conductance seriously overestimates 
the true 
conductance at temperatures higher than the Kondo temperature.
We summarize and draw our conclusions  in 
Section \ref{concsec}.

\section{The xc potential in the constant interaction model}
\label{cimsec}

A popular model of quantum dots is the Constant Interaction Model (CIM) 
described by the Hamiltonian 
\be
\hat{H}_{\rm QD}=\hat{H}_{0}+\hat{H}_{\rm int}
\label{hamil_cim}
\ee
where the non-interacting part of the Hamiltonian is 
$\hat{H}_{0}= \sum_l \e_l \hat{c}_l^{\dagger} \hat{c}_l$. 
The $\e_{l}$ are the 
single-particle eigenvalues of $\hat{H}_{0}$ (the quantum number $l=i\s$ 
comprises an orbital, $i$, and a spin, $\s$, degree of freedom) and, for 
future reference, $\f_{l}$ are the corresponding single-particle orbitals. 
The interaction has the simple form\cite{kat.2001}
\be
\hat{H}_{\rm int}=\frac{1}{2}E_{C}(\hat{N}-N_{0})^{2}
\ee
where $E_{C}$ is the charging energy, 
$\hat{N}=\sum_l \hat{c}_l^{\dagger} \hat{c}_l$ is the operator for the 
total number of electrons and $N_{0}$ is the number of electrons in the 
charge neutral state. Since $\hat{H}_{\rm int}$ depends only on 
$\hat{N}$, and $\hat{N}$ commutes with $\hat{H}_{0}$, the eigenstates of 
$\hat{H}_{0}$ are also eigenstates of $\hat{H}_{\rm QD}$:
\be
\hat{H}_{0}|\Q\ket=E_{0}|\Q\ket \quad\Rightarrow\quad
\hat{H}_{\rm QD}|\Q\ket=E|\Q\ket
\ee
with 
\be
E=E_{0}+\frac{1}{2}E_{C}(N-N_{0})^{2}
\ee
and $\hat{N}|\Q\ket=N|\Q\ket$. 
Therefore the eigenstates of $\hat{H}_{\rm QD}$ with $N$ electrons are Slater 
determinants of $N$ occupied single-particle eigenfunctions 
$\f_{l_{1}},\ldots,\f_{l_{N}}$ and have eigenvalue $E=\e_{l_{1}}+\ldots 
\e_{l_{N}}+\frac{1}{2}E_{C}(N-N_{0})^{2}$.

The KS system of the CIM is described by the KS 
Hamiltonian
\be
\hat{H}_{\rm KS}=\hat{H}_{0}+\int d\blr \;v_{\rm Hxc}(\blr)\hat{n}(\blr)
\ee
where $v_{\rm Hxc}$ is the Hartree-xc potential and $\hat{n}$ is the 
density operator whose integral over all space gives the operator 
$\hat{N}$. The Hartree-xc potential is a universal functional of the 
density, $v_{\rm Hxc}=v_{\rm Hxc}[n]$, and
has the property that for a given temperature $T$ and chemical 
potential $\m$ the equilibrium density of 
$\hat{H}_{\rm QD}$ and $\hat{H}_{\rm KS}$ are identical.
Note that we are using the finite-temperature (or ensemble) DFT as 
formulated in Ref. \cite{mermin}, thus
there is no ambiguity in $v_{\rm Hxc}$ with respect to addition of a 
constant since the number of particles can fluctuate.

Let us consider the quantum dot at zero temperature and denote by 
$E(N)$ the ground-state energy of $\hat{H}_{\rm QD}$ with $N$ electrons.
For a given $\m$ the number of electrons $N$ is  the largest 
integer for which the addition energy
\be
A(N)\equiv E(N)-E(N-1)<\m.
\ee
The ground state of $\hat{H}_{\rm QD}$ with $N$ electrons has, in general, 
degeneracy $d$ and therefore the zero-temperature density can be written 
as
\be
n(\blr)=\frac{1}{d}\sum_{j=1}^{d}\bra\Q_{j}|\hat{n}(\blr)|\Q_{j}\ket
\label{degdens}
\ee
where $|\Q_{j}\ket$ is the $j$-th component of the ground-state 
multiplet. Since every $|\Q_{j}\ket$ is a Slater determinant the 
generic term of the sum in Eq. (\ref{degdens}) has the form
\be
\bra\Q_{j}|\hat{n}(\blr)|\Q_{j}\ket=\sum_{k=1}^{N}|\f^{(j)}_{l_{k}}(\blr)|^{2}
\label{intdens}
\ee
where the $\f^{(j)}_{l_{k}}$'s are the eigenfunctions of the Slater 
determinant $\Q_{j}$.
We now show that for the KS system to 
reproduce the same density of the interacting system the Hartree-xc 
potential has to be uniform in space and depend only on the 
total number of particles, i.e.,
\be
v_{\rm Hxc}[n](\blr)=v_{\rm Hxc}[N].
\label{vhxc}
\ee
Interestingly the fact that $v_{\rm Hxc}$ is uniform in space is not 
an exclusive feature of systems invariant under translations, like the 
electron gas. In the CIM Hamiltonian the one-body part $\hat{H}_{0}$ 
can be any; the uniformity of $v_{\rm Hxc}$ is due to the 
fact that  $\hat{H}_{\rm int}$ depends only on $\hat{N}$. To prove
Eq. (\ref{vhxc}) we simply observe that if $v_{\rm Hxc}$ is independent of $\blr$ then 
the KS eigenfunctions are identical to the 
single-particle eigenfunctions $\f_{l}$ and hence the KS density is 
again given by Eq. (\ref{degdens}) provided that $N$ is the largest 
integer for which the KS eigenvalue
\be
\e_{H(N)}+v_{\rm Hxc}[N]< \m.
\label{ksaddene}
\ee
In Eq. (\ref{ksaddene})  $\e_{H(N)}$ is the noninteracting energy of the HOMO 
level with $N$ electrons; if we 
order the single-particle energies 
according to $\e_{1}\leq \e_{2}\leq \e_{3}\leq \ldots$
then $\e_{H(N)}=\e_{N}$.
It is easy to show that for any {\em real} $N$ the explicit form 
of $v_{\rm Hxc}$ is
\be
v_{\rm Hxc}[N]=A(\bar{N})-\e_{\bar{N}},\quad\quad
\bar{N}\equiv {\rm Int}[N]
\label{vhxc1}
\ee
where $\bar{N}$ is the integer part of $N$. Suppose that there are 
$\bar{N}$ electrons in the quantum dot so that 
$A(\bar{N})<\m<A(\bar{N}+1)$. 
Then $\e_{\bar{N}}+v_{\rm 
Hxc}[\bar{N}]=A(\bar{N})<\m$ whereas $\e_{\bar{N}+1}+v_{\rm 
Hxc}[\bar{N}+1]=A(\bar{N}+1)>\m$; hence $\bar{N}$ is the largest 
integer for which Eq. (\ref{ksaddene}) is fulfilled.
Taking into account the explicit form of the interaction we have
\bea
A(\bar{N})&=&\e_{\bar{N}}+\frac{1}{2}E_{C}(\bar{N}-N_{0})^{2}-\frac{1}{2}E_{C}(\bar{N}-1-N_{0})^{2}
\nn\\
&=&\e_{\bar{N}}+E_{C}(\bar{N}-N_{0}-\frac{1}{2})
\eea
and from Eq. (\ref{vhxc1}) we deduce that
\be
v_{\rm Hxc}[N]=E_{C}\bar{N}-E_{C}(N_{0}+\frac{1}{2}).
\label{isolqd}
\ee
Thus the Hartree-xc potential of the CIM is piece-wise constant with 
discontinuities $E_{C}$ every time $N$ crosses an integer. One
can verify that the size 
of the discontinuities coincides with the xc
part of the derivative discontinuity of the ground states energy, as 
it should be. We mention that there exists generalizations of the CIM 
with $\bar{N}$-dependent discontinuities~\cite{obh.2000}
for which an analytic expression of the Hartree-xc potential can 
still be derived~\cite{ks.2013}.

Next we discuss how to account for the effects of the coupling 
between the quantum dot and the leads. We consider a left ($L$) and a 
right ($R$) lead with Hamiltonian
\be
\hat{H}_{\rm lead}=\sum_{k\a}\e_{k\a}\hat{c}^{\dag}_{k\a}\hat{c}_{k\a}
\ee
where the operators $c_{k\a}$ ($c^{\dag}_{k\a}$) annihilate 
(create) an electron of energy $\e_{k\a}$ in lead 
$\a=L,R$. In the following we assume that the band-width of the leads 
is much larger than any other energy scale and that the quantum-dot 
energies $\e_{l}$ are all well inside the leads band. This is the Wide Band Limit 
Approximation (WBLA).
The contact between the leads and the quantum dot is 
described by the tunneling Hamiltonian
\be
\hat{H}_{\rm T}=\sum_{k\a}\sum_{l}T_{k\a,l}c^{\dag}_{k\a}\hat{c}_{l}+{\rm 
H.c.}
\ee
For a single-level quantum dot, $l=1$, the full Hamiltonian $\hat{H}=H_{\rm 
QD}+\hat{H}_{\rm T}+\hat{H}_{\rm lead}$
reduces to the Hamiltonian of the Anderson model. In this case one 
can show that due to the dot-lead coupling the sharp steps of the 
Hartree-xc potential are smeared. In particular 
the derivative $\de v_{\rm Hxc}/\de N$ at $N=1$ is not infinite but 
instead given by $\p E_{C}^{2}/4\g$~\cite{es.2011}, where $\g$ is the contact-induced 
broadening of the spectral function
\be
\g=2\p \sum_{k\a}|T_{k\a,1}|^{2}\d(\w-\e_{k\a})
\label{litg}
\ee
which is $\w$-independent in the WBLA.
Using the Bethe Ansatz~\cite{wt.1983} it is possible to extract 
the exact form of $v_{\rm Hxc}$ 
by reverse engineering~\cite{blb.2012} and show that it is well 
approximated by (modulo an additive constant)
\be
v_{\rm Hxc}[N]=\frac{1}{\p}E_{C}\arctan\left(\frac{N-1}{W}\right)
\label{fit}
\ee
where $W$ is a parameter that controls the smearing of the step. 
Since $\de v_{\rm Hxc}/\de N\sim E_{C}^{2}/\g$ for $N=1$ 
(see discussion above 
Eq. (\ref{litg})) we have 
$W\sim \g/E_{C}$. The sharp step is recovered in the limit $\g\ra 0$,
which corresponds to the isolated quantum dot. As the only effect of the 
dot-lead coupling is to smear the steps of $v_{\rm Hxc}$ we can easily 
generalize Eq. (\ref{fit}) to multiple-level quantum dots by summing 
over all charged states~\cite{ps.2012}
\be
v_{\rm Hxc}[N]=\frac{1}{\p}E_{C}
\sum_{l}\arctan\left(\frac{N-l}{W(l)}\right) + C,
\label{hxcpotmod}
\ee
where the constant $C$ is chosen such that $v_{\rm Hxc}[0]=0$. 
$W(l)$ accounts for the smearing of the steps and, in general, depends on 
$l$. A 
practical way to estimate the $W(l)$'s consists in constructing the 
broadening matrix
\be
\G_{ll'}=2\p\sum_{k\a}T^{\ast}_{k\a,l}T_{k\a,l'}\d(\w-\e_{k\a})
\label{broad}
\ee
which is $\w$-independent in the WBLA, and then take 
\be
W(l)=\a \G_{ll}/E_{C}
\label{w's}
\ee
with $\a$ a constant of order 1. This choice agrees with the $W$ of 
the Anderson model in the case of a single level. The appealing 
feature of the smeared Hartree-xc potential in Eq. (\ref{hxcpotmod}) 
is that in the limit of weak dot-lead coupling $\G_{ll}\ra 0$, 
hence $W(l)\ra 0$ and $v_{\rm Hxc}$ correctly  reduces (modulo an 
additive constant) to the Hartree-xc potential of the isolated 
quantum dot derived in Eq. (\ref{isolqd}).

Finally we discuss temperature effects. Thermal fluctuations cause a 
further smearing of the steps in $v_{\rm Hxc}$. However, if we are 
interested in temperatures $T<{\rm min}[\{\G_{ll}\}]$  the thermal smearing is 
negligible compared to the smearing induced by the dot-lead 
coupling~\cite{sk.2011,es.2011}. In the remainder of the paper we focus on this 
low temperature regime and use the Hartree-xc potential of Eq. 
(\ref{hxcpotmod}) with the $W$'s from Eq. (\ref{w's}) 
to calculate the density and the KS conductance.

\section{Kohn-Sham density and conductance}
\label{denscondsec}

Due to the fact that $v_{\rm Hxc}$ is uniform in space 
the KS Hamiltonian of the quantum dot is diagonal in 
the basis $\f_{l}$ which diagonalizes $\hat{H}_{0}$:
\be
\hat{H}_{\rm KS}=\sum_{l}(\e_{l}+v_{\rm 
Hxc}[N])\hat{c}^{\dag}_{l}\hat{c}_{l}.
\ee
Let $h_{\rm KS}$ be the single-particle Hamiltonian matrix with 
matrix elements $[h_{\rm KS}]_{ll'}=\d_{ll'}(\e_{l}+v_{\rm Hxc}[N])$ 
in the basis $\f_{l}$. The KS retarded Green's function for the quantum 
dot in contact with the leads can be written as 
\be
\callG_{s}(\w)=\frac{1}{\w-h_{\rm KS}+i\G/2}
\ee
where the broadening matrix has been defined in Eq. (\ref{broad}). To 
deal with out-of-equilibrium situations we split $\G=\G_{L}+\G_{R}$ 
into the sum of the left and right contribution. If we apply a bias 
$V_{\a}$ on lead $\a$ and the system attains a steady state in the 
long-time limit then the steady-state value of the total number of 
electrons in the quantum dot is given by
\be
N=\sum_{\a}\int\frac{d\w}{2\p}f_{\a}(\w)\Tr\left[
\callG_{s}(\w)\G_{\a}\callG_{s}^{\dag}(\w)\right]
\label{selfN}
\ee
where $f_{\a}(\w)=1/(e^{\b(\w-V_{\a}-\m)}+1)$ is the Fermi function at inverse 
temperature $\b=1/T$ and the symbol $\Tr$ signifies a trace over all 
single-particle states of the quantum dot. Since $h_{\rm KS}=h_{\rm 
KS}[N]$ is a function of $N$, Eq. (\ref{selfN}) has to be solved 
self-consistently. With the steady-state value of $N$ the current 
of the KS system can be calculated using the Landauer 
formula~\cite{landauer,meir,jauho}
(the KS electrons are effectively noninteracting)
\be
I=\int\frac{d\w}{2\p}[f_{L}(\w)-f_{R}(\w)]
\Tr\left[\callG_{s}(\w)\G_{L}\callG_{s}^{\dag}(\w)\G_{R}\right]
\label{kscurr}
\ee
where $\callG_{s}$ is evaluated at the self-consistent value of $N$.
The current $I=I(V_{L},V_{R})$ is a function of $V_{L}$ and $V_{R}$ 
and correctly vanishes for $V_{L}=V_{R}$. 
We are interested in the zero-bias  conductance 
which is defined as 
\be
G_{s}=\left.\frac{dI(V/2,-V/2)}{dV}\right|_{V=0}.
\label{gs}
\ee

The procedure outlined above is, in essence, the 
Non-Equilibrium Green's Functions (NEGF) DFT approach to quantum 
transport~\cite{Lang1,Lang2,Taylor1,Taylor2,brand}.
The theorems of DFT guarantee 
that the equilibrium density is exact (provided that the exact $v_{\rm 
Hxc}$ is used) but they do not say anything about out-of-equilibrium 
situations. In this sense, the NEGF-DFT approach may be viewed as an 
empirical approach. 
The rigorous treatment of out-of-equilibrium situations requires to 
formulate the quantum transport problem within Time-Dependent (TD)
DFT~\cite{rg.1984}. This leads to a correction to Eqs. (\ref{selfN}) 
and (\ref{kscurr}) in which 
$V_{\a}\ra V_{\a}+V_{\a,\rm xc}$ where $V_{\a,\rm xc}$ is the 
steady-state value of the TDDFT xc potential in lead 
$\a$~\cite{sa1.2004,sa2.2004,s.2007}.
In the limit of zero bias, $V_{\a,\rm xc}$ vanishes but $dV_{\a,\rm 
xc}/dV$, in general, does 
not~\cite{ewk.2004,sai.2005,kbe.2006,seminario}, and leads to a 
{\em dynamical} correction to the conductance. Adiabatic 
approximations to the TDDFT $v_{\rm Hxc}$, including the exact 
DFT Hartree-xc potential~\cite{tgk.2008,tk.2009},  do not capture this 
correction since  $V_{\a,\rm xc}$ is identically zero in this case. 
Thus the NEGF-DFT approach corresponds to the solution of the TDDFT
equations in the long-time limit {\em and} in the adiabatic approximation. 
This has been nicely confirmed in Refs. \cite{yzcwfn.2011,wyfct.2011}.
In the next section we investigate the performance of the NEGF-DFT 
approach in quantum dots using a DFT 
$v_{\rm Hxc}$, Eq. (\ref{hxcpotmod}), which is essentially 
exact in the weak coupling limit. The analysis will help us to 
understand strengths and limitations of adiabatic approximations in 
quantum transport.

\section{Results and discussions}
\label{ressec}

\begin{figure}[t]
\includegraphics*[width=\linewidth]{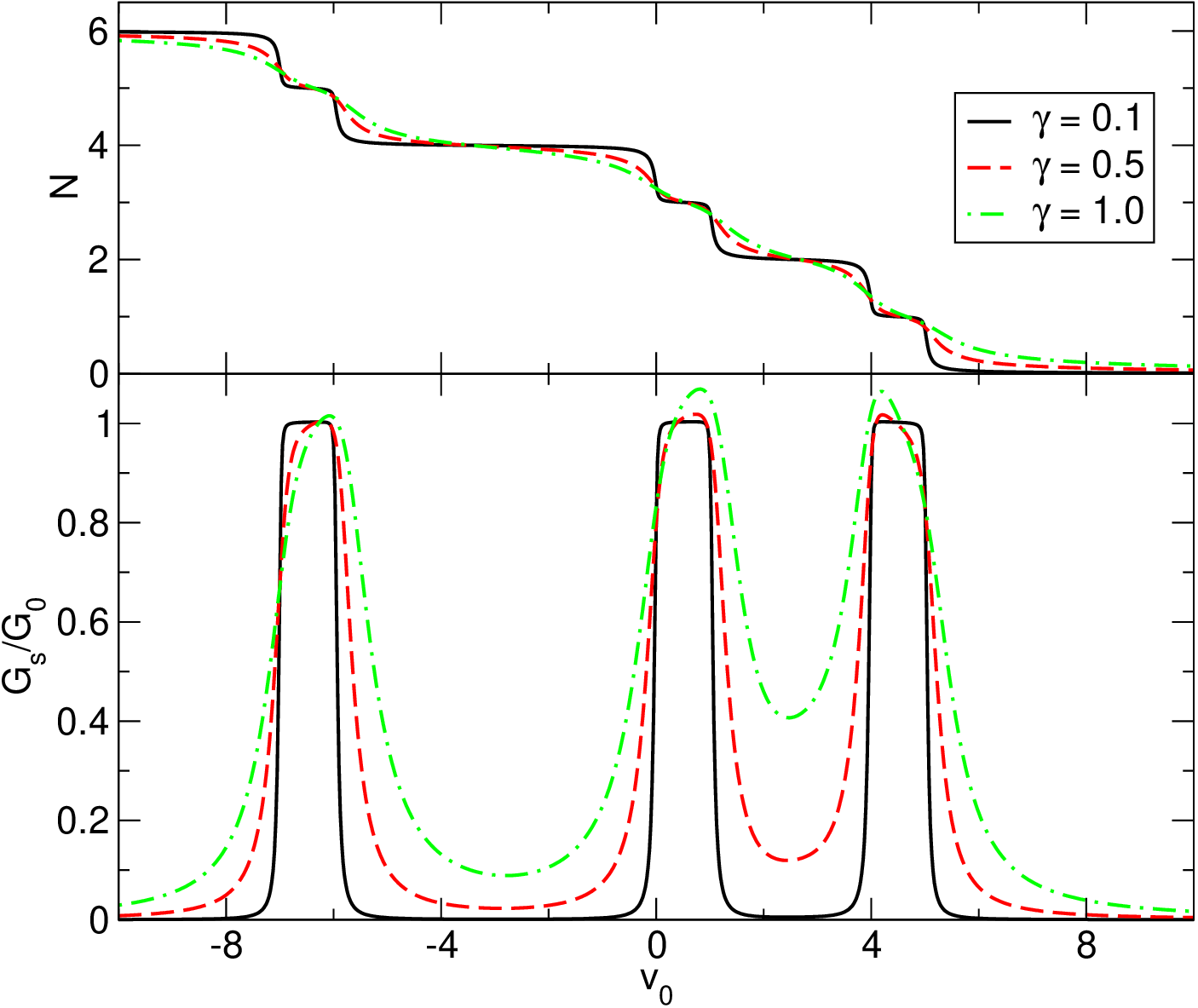}
\caption{Top: Self-consistent solution of Eq. (\ref{selfN}) for the 
total number $N$ of electrons in the quantum dot. Bottom: KS 
conductance $G_{s}$ at zero bias, Eq. (\ref{gs}), in units of the quantum of conductance 
$G_{0}=1/\p$. Both $N$ and $G_{s}$ are studied as function of the 
gate voltage $v_{0}$ for three different values of 
$\g=0.1,\;0.5,\;1.0$.}
\label{fig1}
\end{figure}

Our first example is a three-level quantum dot with each level twice 
degenerate due to spin. The single-particle energies are (in 
arbitrary units) $\e_{l=i\s}=\e_{i}^{0}+v_{0}$ with $v_{0}$ the gate 
voltage,
$\e_{1}^{0}=-5$, $\e_{2}^{0}=-3$ and $\e_{3}^{0}=2$, the 
charging energy $E_{C}=1$, and the broadening matrices $\G_{\a}$ are  diagonal 
with diagonal entries equal to $\g/2$. We take $\m=0$ and work at 
temperature $T=0.001\ll\g$, so that the thermal broadening can be 
discarded and Eq. (\ref{hxcpotmod}) is a good approximation for 
$v_{\rm Hxc}$. Since the broadening is the same for all levels we 
choose the smearing parameter $W(l)=0.16 \g/E_{C}$ independent of 
$l$. In Fig. \ref{fig1} (top) we show the self-consistent 
solution of Eq. (\ref{selfN}) versus gate voltage $v_{0}$ for three different values of 
$\g=0.1,\;0.5,\;1.0$. The discontinuity in $v_{\rm Hxc}$ is crucial to 
pin the levels to the chemical potential and hence to give rise to the 
Coulomb blockade ladder. For $v_{0}>5$ the quantum dot is empty since 
all levels are above $\m$. For $v_{0}=5$ one electron enters the 
dot and sits on the first energy level. Due to the fact that the addition of a second electron costs a 
charging energy $E_{C}=1$ it is not until $v_{0}$ lowers to 4 
that a second electron enters. At this point  the quantum dot is in a spin 
singlet with two electrons occuping the first level. 
The presence of two electrons shifts the unoccupied levels to 
$\e_{i\s}+v_{\rm Hxc}[2]=\e_{i\s}+2E_{C}$. Therefore for a 
third electron to enter the  dot $v_{0}$ has to lower to 
$v_{0}=-\e_{2}^{0}-2E_{C}=3-2=1$. These energetic considerations explain the 
curve $N$ versus $v_{0}$ from $N=0$, when the quantum dot is empty,
to $N=6$, when the quantum dot is totally filled. 
The message to take home is that the plateaus of the $N$-$v$ 
curve have length $E_{C}$ if the number of electrons is odd and 
$E_{C}+\e^{0}_{i+1}-\e^{0}_{i}$ if the number of electrons is even.

Once $N=N(v_{0})$ is known we can proceed with the calculation of the 
conductance. In Fig. \ref{fig1} (bottom) we show the zero bias 
KS conductance of Eq. (\ref{gs}). 
Interestingly $G_{s}$ exhibits a plateau whenever $N$ 
is odd, and this behavior is the same as that of the true conductance~\cite{onh.2005,no.2006,yss.1998}. 
The plateau is indeed a consequence of
the fact that at low temperatures the true spectral 
function exhibits a sharp Kondo peak at the chemical 
potential. 
The results of  Fig. \ref{fig1} constitute a generalization to multiple levels of the 
findings in Refs. ~\cite{sk.2011,blb.2012,tse.2012} and can be 
explained using the Friedel sum rule~\cite{langreth,mera1,mera2}. We can conclude 
that the 
NEGF-DFT approach, even though based on an adiabatic approximation, 
is capable to capture strong correlation effects provided that
the exact (discontinuous)  Hartree-xc potential is 
used.

\begin{figure}[t]
\includegraphics*[width=\linewidth]{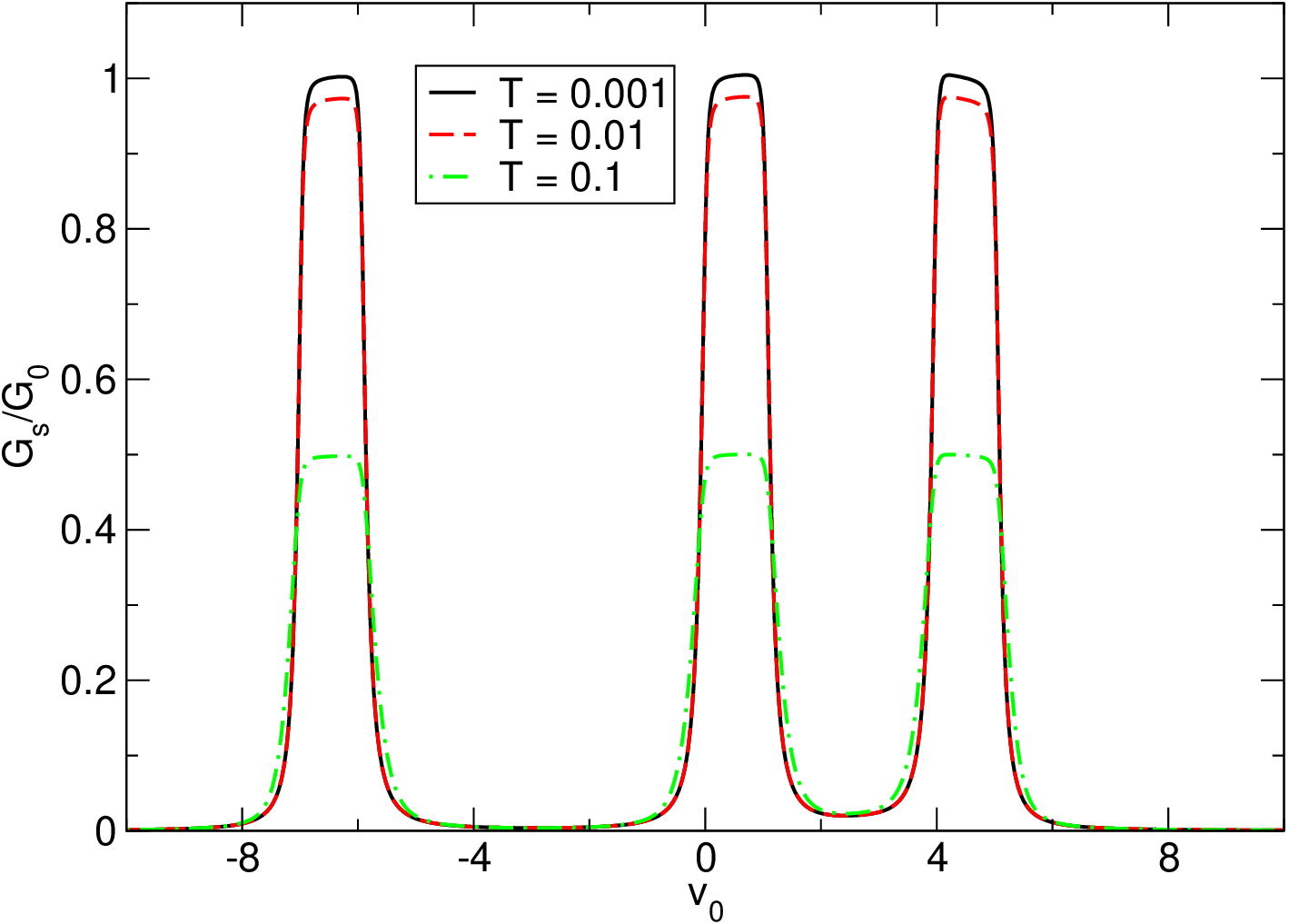}
\caption{KS conductance $G_{s}$ at zero bias, Eq. (\ref{gs}), in units of the quantum of conductance 
$G_{0}=1/\p$ versus the gate voltage $v_{0}$ for $\g=0.2$ and three different values of 
the temperature $T=0.001,\;0.01,\;0.1$.}
\label{fig2}
\end{figure}

Unfortunately the agreement between the KS conductance and the true 
conductance deteriorates with increasing temperature. In Fig. 
\ref{fig2} we display $G_{s}$ versus gate voltage for 
$\g=0.2$ and three different values of  
the temperature $T=0.001,\;0.01,\;0.1$.
For temperatures $T$ larger than the Kondo temperature $T_{K}\leq 
\sqrt{2E_{C}\g}\;e^{-\p E_{C}/8\g}\simeq 0.089$ the middle part of the plateau should be 
suppressed leaving two Coulomb blockade peaks~\cite{IzumidaSakai:05,Costi:00}. Instead we see that
the entire plateau is suppressed and there is no signature of the Coulomb 
blockade peaks. 
As discussed at the end of Section \ref{cimsec}, 
the DFT Hartree-xc potential is weakly dependent on temperature 
for $T$ smaller than the level broadenings, and therefore our approximation is  
accurate in this temperature range. Thus, we infer that at finite temperature the adiabatic 
approximation (which, we recall, is an intrinsic feature of the NEGF+DFT approach) 
is not able to reproduce the true conductance in the Coulomb blockade 
regime. In this temperature range the dynamical xc correction of TDDFT 
is essential~\cite{sk.2011}. This  correction involves 
the TDDFT kernel with coordinates deep inside the 
leads~\cite{sk.2011} and, therefore, is rather difficult to estimate.
In fact, the Coulomb blockade regime has 
remained out of reach until very recently. In Ref.~\cite{ks.2013} 
we solved the problem and provided a comprehensive (TD)DFT picture of 
the Coulomb blockade regime 
without breaking the spin symmetry.

\begin{figure}[t]
\includegraphics*[width=\linewidth]{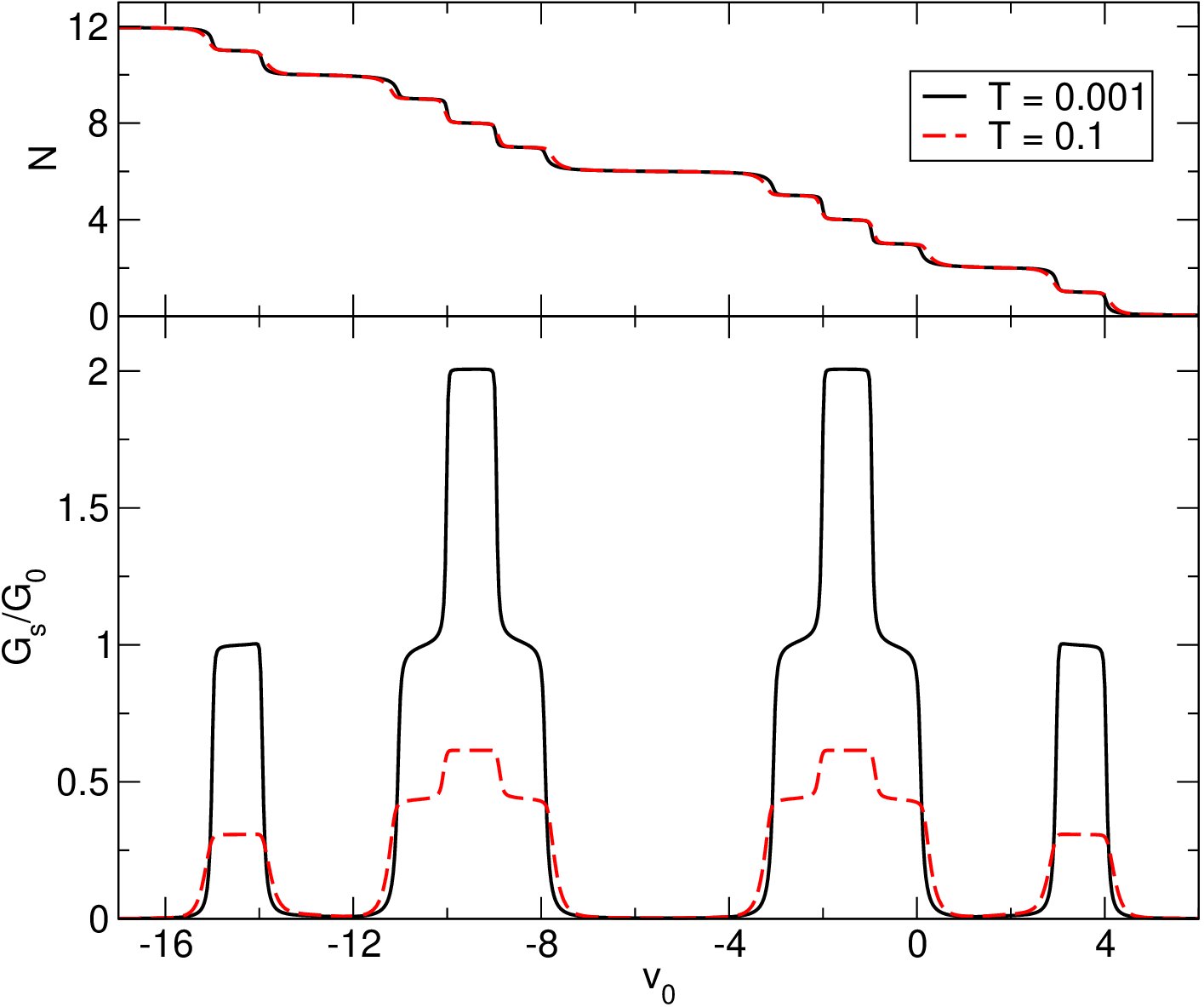}
\caption{Top: Self-consistent solution of Eq. (\ref{selfN}) for the 
total number $N$ of electrons in the quantum dot. Bottom: KS 
conductance $G_{s}$ at zero bias, Eq. (\ref{gs}), in units of the quantum of conductance 
$G_{0}=1/\p$. Both $N$ and $G_{s}$ are studied as function of the 
gate voltage $v_{0}$ for two different values of temperature
$T=0.001,\;0.1$.}
\label{fig3}
\end{figure}

The inadequacy of the NEGF+DFT approach at finite temperature was 
somehow expected since the relation~\cite{langreth} between 
the phase-shift (hence the conductance) and the density does not 
hold any longer. Independently of temperature, no such relation exists for quantum dots with 
degenerate levels {\em unless the broadening matrices are diagonal}.
In these cases we expect that at sufficiently low temperature the KS conductance is 
again in qualitative good agreement with the true conductance. 

Let us consider a six-level quantum dot with 
single-particle energies $\e_{l=i\s}=\e_{i}^{0}+v_{0}$ with $v_{0}$ the 
gate voltage and $\e_{1}^{0}=-4$, $\e_{2}^{0}=\e_{3}^{0}=-2$, 
$\e_{4}^{0}=\e_{5}^{0}=2$, $\e_{6}^{0}=4$. As in the previous 
example each level is two-fold degenerate due to spin. We take 
the charging energy $E_{C}=1$, the broadening matrices diagonal, 
$[\G_{\a}]_{i\s,j\s'}=\d_{ij}\d_{\s\s'}\g/2$, with $\g=0.1$, and hence 
the smearing parameter $W(l)=0.16 \g/E_{C}$ independent of 
$l$. The chemical potential 
is set to $\m=0$. In Fig. \ref{fig3} we show the number of electrons 
$N$ (top) and KS conductance $G_{s}$ (bottom) versus gate voltage 
$v_{0}$ for two different temperatures $T=0.001,\;0.1$. According to 
our parameters when $v_{0}\gtrsim 0$ we have $\e_{2}^{0}+v_{0}+v_{\rm 
Hxc}[2]=-2+v_{0}+2E_{C}\gtrsim 0$ and therefore the dot has two electrons occupying 
the lowest energy level. For $v_{0}\lesssim 0$ a third 
electron enters the dot. At this stage the levels $\e^{0}_{2,3}+v_{0}+v_{\rm Hxc}[3]$ 
are very close to $\m$ but not exactly pinned to $\m$ since the 
occupancy is $1/2$  for each of them.
The KS conductance is the sum of the 
noninteracting conductances of two single-level models with occupancy 
$1/2$ since the $\G_{\a}$ are diagonal. For occupancy $1/2$
the single-level 
conductance is $G_{0}/2$  and hence 
$G_{s}=2G_{0}/2=G_{0}$, in agreement with Fig. \ref{fig3}. 
Suppose now to lower $v_{0}$ further until the value
$v_{0}\simeq -E_{C}$, when a fourth electron enters the dot.
With four electrons the levels $\e^{0}_{2,3}+v_{0}+v_{\rm Hxc}[4]$ are exactly 
pinned to $\m$ since the occupancy is $1$  for each of them.
For occupancy 1 the single-level 
conductance is $G_{0}$ and hence $G_{s}=2G_{0}$. This explains the 
second plateau at $2G_{0}$ for $-1\lesssim v_{0}\lesssim 0$. 
With similar arguments we can understand all remaining 
plateaus. This behavior of $G_{s}$ is qualitatively similar to that of 
the true conductance~\cite{yss.1998,lyfmt.1999}.
The effect of increasing temperature on $G_{s}$ is to lower the 
plateaus. The true conductance, instead,  
develops distinct Coulomb blockade peaks~\cite{yss.1998,lyfmt.1999}. 
We conclude that the NEGF-DFT approach accounts for degeneracy 
effects reasonably well (for diagonal $\G$-matrices)
whereas it needs to be considerably revised in order to include 
finite temperature effects (Coulomb blockade regime)~\cite{ks.2013}.

\begin{figure}[t]
\includegraphics*[width=\linewidth]{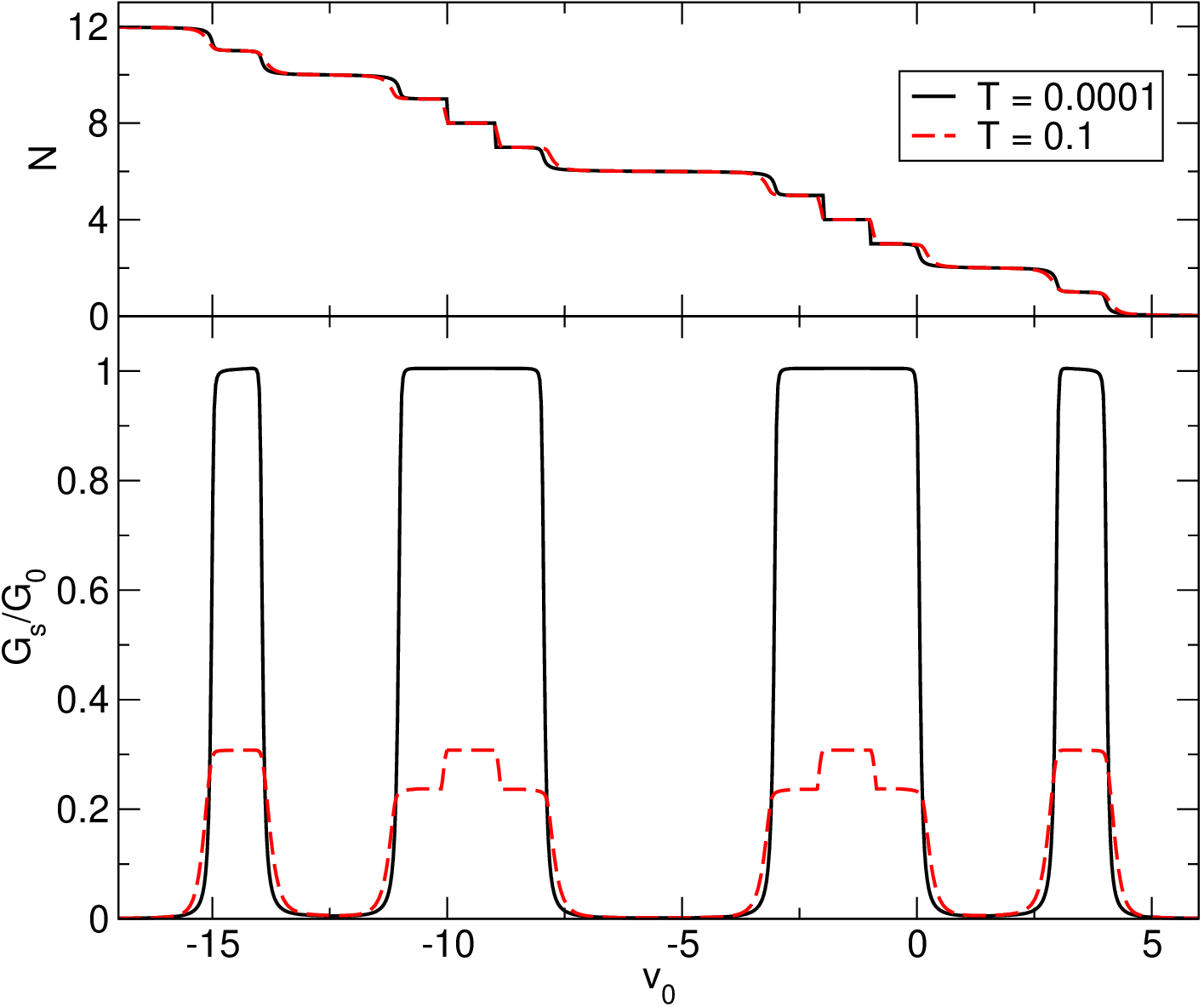}
\caption{Same system as in Fig. \ref{fig3} except for 
$[\G_{\a}]_{3\s,3\s}=[\G_{\a}]_{5\s,5\s}=0$. The self-consistent 
solution for $N$ (top) has been obtained by modifying the Hartree-xc 
potential as described in the main text.}
\label{fig4}
\end{figure}

We also studied a variant of the previous six-levels quantum dot in 
which one of the degenerate levels in each multiplet, say $\e_{3\s}$ 
and $\e_{5\s}$, do not hybridize with the leads and hence 
$[\G_{\a}]_{3\s,3\s}=[\G_{\a}]_{5\s,5\s}=0$. 
Consider the dot with two electrons in the lowest energy level $\e_{1\s}$. By lowering 
$v_{0}$ a third electron enters the dot. Unlike the previous case 
this electron does not equally distribute between levels  $\e_{2\s}$ 
and  $\e_{3\s}$ but  occupies only level  $\e_{2\s}$.
Indeed $\e_{2\s}$ hybridizes with the leads and hence it gets half 
populated when $\e_{2\s}=0+\d$ with $\d$ a small positive constant. 
Since any $\d>0$ is larger than the broadening of level 
$\e_{3\s}$, this level remains empty.  
By lowering $v_{0}$ further we eventually 
start occupying level 
$\e_{3\s}$. The population of this level occurs very rapidly 
since it is not coupled to the leads. From 
these considerations we see that the form of $v_{\rm Hxc}$ in Eq. 
(\ref{fit}) needs to be modified. For instance, according to the 
previous argument, for $2<N<3$ the smearing parameter $W(3)$
is finite but for $3<N<4$ the same smearing parameter  
vanishes. Therefore we introduce $N$-dependent smearing parameters 
$W(l,N)$ in Eq. (\ref{fit}) according to
\bea
W(3,N)&=&\th(3-N)W\nn\\
W(4,N)&=&0\nn\\
W(5,N)&=&\th(N-5)W\nn\\
W(7,N)&=&\th(7-N)W\nn\\
W(8,N)&=&0\nn\\
W(9,N)&=&\th(N-9)W\nn
\eea
and $W(1)=W(2)=W(6)=W(10)=W(11)=W(12)=W=0.16\g/E_{C}$. The 
self-consistent solution of Eq. (\ref{selfN}) is shown in Fig. 
\ref{fig4} (top). The modified Hartree-xc potential correctly reproduces 
the trend of $N$ versus $v_{0}$ deduced from physical arguments. The 
transitions $0\to 1\to 2\to 3$ electrons are smooth whereas the 
transitions $3\to 4\to 5$ are sharp. The same behavior is 
observed for the other pair of degenerate levels. The corresponding KS 
conductance is displayed in Fig. \ref{fig4} (bottom). The main 
difference with the previous case is that the plateau at $2G_{0}$ 
lowers to $G_{0}$. This can be understood as follows. With 3 
electrons in the quantum dot the level $\e_{2\s}$ is half filled while 
level $\e_{3\s}$ is empty. Since level $\e_{2\s}$ is coupled to the 
leads the conductance is $G_{s}=G_{0}$. With 4 and 5 electrons level 
$\e_{2\s}$ remains half filled and level $\e_{3\s}$ is half-filled 
and totally filled respectively. This level, however, is not coupled 
to the leads and hence does not contribute to the conductance which
remains equal to $G_{0}$. With similar energetic considerations we 
can explain all remaining plateaus. With increasing temperature the 
thermal broadening of the Fermi function allows for a slight 
distribution of electrons. When a third electron enters the dot we 
have little occupancy of level $\e_{3\s}$ as well. 
Thus level  $\e_{2\s}$ is not exactly half filled and the conductance is 
lower than that with 4 electrons when level $\e_{2\s}$ is instead 
half filled. 

\section{Conclusions and outlooks}
\label{concsec}

We generalized the findings of Ref. \cite{sk.2011} to multiple 
levels, relevant for quantum dots or molecular junctions. 
At temperatures below the Kondo temperature and for 
nondegenerate levels the NEGF-DFT approach is very accurate provided 
that the exact DFT Hartree-xc potential is used. The success of 
NEGF-DFT has its origin in the fulfillment of the Friedel 
sum rule.  At finite temperature the Friedel 
sum rule is of no use to relate the conductance to the density and we 
indeed find that the KS and true conductances can 
differ quite substantially. 
Another situation for which, in general, we cannot 
use the Friedel sum rule is that of degenerate levels. 
An exception is that of diagonal $\G$ matrices. In this case the 
NEGF-DFT approach is again exact provided that the exact DFT Hartree-xc 
potential is employed. For the CIM we 
constructed an approximation to $v_{\rm Hxc}$ which becomes exact in the 
limit of weak coupling. Our results for the (KS) conductance are in good 
qualitative agreement with numerical renormalization group 
results~\cite{yss.1998} and many-body results~\cite{lyfmt.1999}.
In all other cases one has to go beyond the NEGF-DFT approach and 
include dynamical xc corrections in the 
theory. This can be done either by finding suitable approximations to 
the TDDFT kernel and then solving the linear response equations or by 
including memory effects in the TDDFT Hartree-xc 
potential~\cite{dbg.1997,neepa1,vbdvls.2005,t.2007,neepa2} and then
performing time propagations. There has been 
considerable progress in the implementation of {\em ab initio} 
propagation schemes for open 
systems~\cite{bsn.2004,ksarg.2005,zwymc.2007,sbhdv.2007,spc.2010,yzcwfn.2011,wyfct.2011,zww.2012,zxw.2012,zcc.2013} but, 
at present, the results are 
limited to adiabatic Hartree-xc potentials. The  development of 
propagation algorithms for open systems with nonadiabatic Hartree-xc 
potentials represents one of the future challenges in 
quantum transport.

\begin{acknowledgement}
S.K. acknowledges funding by the ``Grupos Consolidados UPV/EHU del
Gobierno Vasco'' (IT-578-13). G.S.  acknowledges funding by MIUR FIRB 
grant No. RBFR12SW0J.
We also acknowledge financial support through 
travel grants (Psi-K2 4665 and 3962 (G.S.), and Psi-K2 5332 (S.K.)) of the 
European Science Foundation (ESF).
\end{acknowledgement}

\end{document}

%% file: ourmacros.tex
\newcommand{\be}{\begin{equation}}
\newcommand{\ee}{\end{equation}}
\newcommand{\bea}{\begin{eqnarray}}
\newcommand{\eea}{\end{eqnarray}}

\def\a{\alpha}
\def\b{\beta}
\def\g{\gamma}
\def\G{\Gamma}
\def\d{\delta}

\def\e{\epsilon}

\def\th{\theta}

\def\m{\mu}

\def\p{\pi}

\def\s{\sigma}

\def\f{\phi}

\def\w{\omega}

\def\Q{\Psi}

\def\blr{{\mathbf r}}

\def\callG{\mbox{$\mathcal{G}$}}

\def\ra{\rightarrow}

\def\de{\partial}

\def\bra{\langle}
\def\ket{\rangle}

\def\Tr{{\rm Tr}}

\def\1op{\hat{\mathbbm{1}}}
\def\nn{\nonumber}

%% file: pssb-Kondo_v2.bbl
\begin{thebibliography}{[1]}
    
\bibitem{loc.2002}
N. A. Lima, L. N. Oliveira and K. Capelle, 
Europhys. Lett. {\bf 60}, 601 (2002).

\bibitem{lsoc.2003}
N. A. Lima, M. F. Silva, L. N. Oliveira and K. Capelle,
Phys. Rev. Lett. {\bf 90}, 146402 (2003).

\bibitem{mscy.2009}
P. Mori-Sanchez, A. J. Cohen and W. Yang, 
Phys. Rev. Lett. {\bf 102}, 066403 (2009).

\bibitem{mgg.2012}
F. Malet and P. Gori-Giorgi, 
Phys. Rev. Lett. {\bf 109}, 246402 (2012).

\bibitem{mmcrg.2013}
F. Malet, A. Mirtschink, J. C. Cremon, S. M. Reimann and P. 
Gori-Giorgi,
Phys. Rev. B {\bf 87}, 115146 (2013).

\bibitem{xctk.2012}
Gao Xianlong, A-Hai Chen, I. V. Tokatly and S. Kurth,
Phys. Rev. B {\bf 86}, 235139 (2012).

\bibitem{lyb.2012}
J. Lorenzana, Z.-J. Ying and V. Brosco,
Phys. Rev. B {\bf 86}, 075131 (2012).

\bibitem{kk.2013}
E. Kraisler and L. Kronik, 
Phys. Rev. Lett. {\bf 110}, 126403 (2013).

\bibitem{v.2008} 
C. Verdozzi, 
Phys. Rev. Lett. {\bf 101}, 166401 (2008).

\bibitem{kskvg.2010}
S. Kurth, G. Stefanucci, E. Khosravi, C. Verdozzi and
E.~K.~U. Gross,
Phys. Rev. Lett. {\bf 104}, 236801 (2010).

\bibitem{hk.2012}
D. Hofmann and S. K\"ummel, 
Phys. Rev. B {\bf 86}, 201109 (2012).

\bibitem{efrm.2012}
P. Elliott, J. I. Fuks, A. Rubio and N. T. Maitra,
Phys. Rev. Lett. {\bf 109}, 266404 (2012).

\bibitem{nrvl.2013}
S. E. B. Nielsen, M. Ruggenthaler and R. van Leeuwen,
EPL {\bf 101}, 33001 (2013).  

\bibitem{pplb.1982}
J. P. Perdew, R. G. Parr, M. Levy, and J. L. Balduz, 
Phys. Rev. Lett. {\bf 49}, 1691 (982).

\bibitem{sk.2011}
G. Stefanucci and S. Kurth,
Phys. Rev. Lett. {\bf 107}, 216401 (2011).

\bibitem{blb.2012}
J. P. Bergfield, Z.-F. Liu and K. Burke,
Phys. Rev. Lett. {\bf 108}, 066801 (2012).

\bibitem{tse.2012}
P. Tr\"oster, P Schmitteckert and F. Evers,
Phys. Rev. B {\bf 85}, 115409 (2012).

\bibitem{kat.2001}
L. P. Kouwenhoven, D. G. Austing and S. Tarucha, 
Rep. Prog. Phys. {\bf 64}, 701 (2001).

\bibitem{mermin}
N. D. Mermin, 
Phys. Rev. {\bf 137}, A1441 (1965).

\bibitem{obh.2000}
Y. Oreg, K. Byczuk and B. I. Halperin, 
Phys. Rev. Lett. {\bf 85}, 365 (2000).

\bibitem{ks.2013}
S. Kurth and G. Stefanucci, Phys. Rev. Lett. {\bf 111}, 030601 (2013).

\bibitem{es.2011}
F. Evers and P. Schmitteckert, 
Phys. Chem. Chem. Phys. {\bf 13}, 14 417 (2011).

\bibitem{wt.1983} 
P. B. Wiegmann and A. M. Tsvelick, 
J. Phys. C: Solid State Phys. {\bf 16}, 2281 (1983).

\bibitem{ps.2012}
E. Perfetto and G. Stefanucci, 
Phys Rev B {\bf 86}, 081409 (2012).

\bibitem{landauer}
R. Landauer, 
IBM J. Res. Dev. {\bf 1},  233  (1957).

\bibitem{meir}
Y. Meir and N.~S. Wingreen,
Phys. Rev. Lett. {\bf 68},  2512  (1992).

\bibitem{jauho}
A.-P. Jauho, N.~S. Wingreen and Y. Meir,
Phys. Rev. B {\bf 50},  5528 (1994).

\bibitem{Lang1}
N.~D. Lang, 
Phys. Rev. B {\bf 52},  5335  (1995).

\bibitem{Lang2}
N.~D. Lang and P. Avouris, 
Phys. Rev. Lett. {\bf 81},  3515  (1998).

\bibitem{Taylor1}
J. Taylor, H. Guo and J. Wang, 
Phys. Rev. B {\bf 63},  121104  (2001).

\bibitem{Taylor2}
J. Taylor, H. Guo and J. Wang, 
Phys. Rev. B {\bf 63},  245407  (2001).

\bibitem{brand}
M. Brandbyge,  J.-L. Mozos, P. Ordej\'on, J. Taylor and K. Stokbro,
Phys. Rev. B {\bf 65},  165401  (2002).

\bibitem{rg.1984}
E. Runge and E.~K.~U. Gross, 
Phys. Rev. Lett. {\bf 52},  997  (1984).

\bibitem{sa1.2004}
G. Stefanucci and C.-O. Almbladh,
Phys. Rev. B {\bf 69}, 195318 (2004).

\bibitem{sa2.2004}
G. Stefanucci and C.-O. Almbladh,
Europhys. Lett. {\bf 67}, 14 (2004).

\bibitem{s.2007}
G. Stefanucci, 
Phys. Rev. B {\bf 75}, 195115 (2007). 

\bibitem{ewk.2004}
F. Evers, F. Weigend and M. Koentopp,
Phys. Rev. B {\bf 69}, 235411 (2004).

\bibitem{sai.2005}
N. Sai, M. Zwolak, G. Vignale and M. Di Ventra,
Phys. Rev. Lett. {\bf 94}, 186810 (2005).

\bibitem{kbe.2006}
M. Koentopp, K. Burke and F. Evers, 
Phys. Rev. B 73, 121403 (2006).

\bibitem{seminario}
G. Stefanucci, S. Kurth, E.~K.~U. Gross and A. Rubio, 
Time-dependent transport phenomena, in: J. M. Seminario (Ed.), 
{\em Molecular and nano electronics: analysis, design and simulation}, 
Vol. 17 of Elsevier Series on Theoretical and Computational
Chemistry, Elsevier, 2007, p. 247.

\bibitem{tgk.2008}
M. Thiele, E.~K.~U. Gross and S. K\"ummel,
Phys. Rev. Lett. {\bf 100}, 153004 (2008).

\bibitem{tk.2009}
M. Thiele and S. K\"ummel,
Phys. Rev. A {\bf 79}, 052503 (2009).

\bibitem{yzcwfn.2011}
C. Yam, X. Zheng, G. H. Chen, Y. Wang, T. Frauenheim and T. A. Niehaus,
Phys. Rev. B {\bf 83}, 245448 (2011).

\bibitem{wyfct.2011}
Y. Wang, C. Yam, T. Frauenheim, G. H. Chen and T. A. Niehaus,
Chem. Phys. {\bf 391}, 69 (2011).

\bibitem{onh.2005}
A. Oguri, Y. Nisikawa and A. C. Hewson,
J. Phys. Soc. Jpn. {\bf 74}, 2554 (2005).

\bibitem{no.2006}
Y. Nisikawa and A. Oguri,
Phys. Rev. B {\bf 73}, 125108 (2006).

\bibitem{yss.1998}
W. Izumida, O. Sakai and Y. Shimizu 
J. Phys. Soc. Jpn. {\bf 67},  2444  (1998).

\bibitem{langreth}
D. C. Langreth, 
Phys. Rev. {\bf 150}, 516 (1966).

\bibitem{mera1}
H. Mera, K. Kaasbjerg, Y. M. Niquet, and G. Stefanucci,
Phys. Rev. B {\bf 81}, 035110 (2010).

\bibitem{mera2}
H. Mera and Y. M. Niquet, 
Phys. Rev. Lett. {\bf 105}, 216408 (2010).

\bibitem{IzumidaSakai:05}
W. Izumida and O. Sakai, 
J. Phys. Soc. Jpn. {\bf 74},  103  (2005).

\bibitem{Costi:00}
T.~A. Costi, 
Phys. Rev. Lett. {\bf 85},  1504  (2000).

\bibitem{lyfmt.1999}
A. Levy Yeyati, F. Flores and A. Mart\`in-Rodero,
Phys. Rev. Lett. {\bf 83}, 600 (1999).

\bibitem{dbg.1997}
J. F. Dobson, M. J. B\"unner and E. K. U. Gross,
Phys. Rev. Lett. {\bf 79}, 1905 (1997).

\bibitem{neepa1}
N. T. Maitra, K. Burke and C. Woodward,
Phys. Rev. Lett. {\bf 89}, 023002 (2002)

\bibitem{vbdvls.2005}
U. von Barth, N. E. Dahlen, R. van Leeuwen and G. Stefanucci,
Phys. Rev. B {\bf 72}, 235109 (2005).

\bibitem{t.2007}
I. V. Tokatly,
Phys. Rev. B {\bf 75}, 125105 (2007).

\bibitem{neepa2}
N. T. Maitra,
Lectures Notes in Physics {\bf 837}, 167 (2012).

\bibitem{bsn.2004}
R. Baer, T. Seideman, S. Ilani, and D. Neuhauser, 
J. Chem. Phys. {\bf 120}, 3387 (2004).

\bibitem{ksarg.2005}
S. Kurth, G. Stefanucci, C.-O. Almbladh, A. Rubio and E.~K.~U. Gross, 
Phys. Rev. B {\bf 72}, 035308 (2005).

\bibitem{zwymc.2007}
X. Zheng, F. Wang, C. Y. Yam, Y. Mo and G. H. Chen,
Phys. Rev. B {\bf 75}, 195127 (2007).

\bibitem{sbhdv.2007}
N. Sai, N. Bushong, R. Hatcher, and M. Di Ventra, 
Phys. Rev. B {\bf 75}, 115410 (2007).

\bibitem{spc.2010}
G. Stefanucci, E. Perfetto and M. Cini,
Phys. Rev. B {\bf 81}, 115446 (2010).

\bibitem{zww.2012}
L. Zhang, B. Wang and Jian Wang,
Phys. Rev. B {\bf 86}, 165431 (2012).

\bibitem{zxw.2012}
L. Zhang, Y. Xing and J. Wang,
Phys. Rev. B {\bf 86}, 155438 (2012).

\bibitem{zcc.2013}
Y. Zhang, S. Chen and G. H. Chen,
Phys. Rev. B {\bf 87}, 085110 (2013).


\end{thebibliography}
